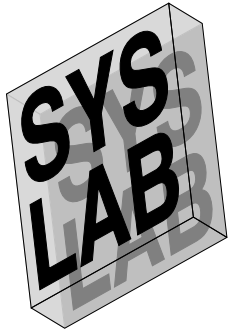

# A stream-based mathematical model for distributed information processing systems
## - SysLab system model -

Cornel Klein, Bernhard Rumpe, Manfred Broy

Institut für Informatik, Technische Universität München

80333 München, Germany

e-mail: (broy|klein|rumpe)@informatik.tu-muenchen.de

**Abstract**

In the SysLab-project, we develop a software engineering method based on a mathematical foundation. The SysLab system model serves as an abstract mathematical model for information systems and their components. It is used to formalize the semantics of all used description techniques, such as object diagrams, state automata, sequence charts or data-flow diagrams. Based on the requirements for such a reference model, we define the system model including its different views and their relationships.

---

[0] This paper originated in the SysLab project, which is supported by the DFG under the Leibnizprogramme and by Siemens-Nixdorf.

[0] It is published in the proceedings of FMOODS'96 (Formal Methods for Open Object-based Distributed Systems) Workshop as report ENST 96S001 by ENST France Telecom, pages 323-338. Editors: Elie Najm and Jean-Bernard Stefani.

# 1 Introduction

Methods for systems and software development, like OMT [RBP+92], Fusion [CAB+94], and GRAPES [Hel90], model a system at different abstraction levels and under different views. Within the process of modeling they provide description techniques like entity-/relationship-diagrams and their object-oriented extensions, state automata, sequence charts or data-flow diagrams. A critical point of existing commercial methods is imprecision of the semantic description. The definition of the description techniques as well as the relationships between different description techniques of a method is usually only given informally. A lot of problems during the application of the methods exist, which are caused by the ambiguous and vague interpretation of the semantics of the used modeling concepts:

- the communication between the persons involved in the project is more difficult, because of ambiguities arising from informal semantic descriptions,

- it is impossible to define formal relationships between different description levels and to define rules to transfer information between two description levels,

- a solid basis for tool support is missing,

- even in one description level there is a lack of clarity concerning the consistency and completeness of a set of documents. Issues concerning "consistency" and "completeness" can only be tackled informal.

As a consequence, tool systems for the support of methods ("CASE-Tools") often do not cause the expected gain in productivity: The information which can be acquired by the use of methods is, because of the deficient semantic foundation of the methods, not very evident. As a result of this, the functionality of tools is mostly restricted to document editing- and managing functions.

Recently, various approaches for formalizing methods of systems and software development were given. Well known are the so-called "meta-models", originating in the context of tool integration, (see [CDI92], [Tho89] and [HL93]). However, by this "models" almost only the abstract syntax of the description techniques is captured. An overview of several projects concerning the integration of structured methods with techniques of formal specification can be found in [SFD92]. In [Hus94], the British standard method SSADM [AG90] is formalized using the algebraic specification language SPECTRUM [BFG+93]. The work of Hussmann goes beyond the approaches described in [SFD92]. Hussmann states a mathematical model of the information systems modeled by SSADM to which he relates the different description techniques which occur in in the method. This approach offers a complete analysis of the semantics of the SSADM-description techniques and their relationships, the definition of conditions for consistency and completeness of a set of description techniques, and a simple basis for obtaining prototypes by functional programs.

## 1.1 The role of the system model in SysLab

The SysLab-project aims at developing a practicable method for system- and software development, that is scientifically founded, and that does not show the above-mentioned disadvantages due to the lack of a semantic foundation. Moreover, in SysLab a prototype of a tool system should be created. The formalization should not end in itself, but it should provide the semantic basis for the check for consistency of the concepts. The semantic foundation is achieved by the usage of a uniform mathematical system model for SysLab. This abstract mathematical model of information processing systems serves for relating to it all description techniques used in SysLab, such as object diagrams, state diagrams, data-flow diagrams, etc., and all transformation rules for the transformation of documents. Each document, such as an object diagram, is regarded as a proposition over the mathematical system model.

The formalization of description techniques leads primarily to a deeper comprehension of the meaning of the descriptions, the aspects on which statements are given, and their inter-relations. Therefore description techniques can be used more objectively. Furthermore it is possible to state conditions for consistency and completeness of a set of description documents, and to define and to analyze relationships between description documents of different abstraction levels. Finally formalization is an important mile-stone on the way to a more effective tool support of methods, because semantic-preserving transformations between different description techniques are feasible which finally result in executable code. Moreover a flexible application of formal techniques, which is necessary in safety-critical applications, is possible.

## 1.2 Requirements on the system model

It is the aim of this paper to provide a common basis for all people involved in the SysLab-project concerning the notion of a system used and the definition of the semantic of the various description techniques. Therefore, the system model has to cover all phases and all description techniques of the SysLab method, and it may not be restricted to a certain class of information processing systems, such as commercial information systems. From that results the requirement to develop a system model which is as *general as possible*.

On the other hand, it should be easy to define a semantics based on the system model for the description techniques to be developed. This leads to the requirement that the system model has to be tailored for the description techniques we are aiming at. This means for instance that we are aiming at a model supporting the dynamic creation and deletion of components ("objects").

The basic assumption with respect to the structure of information processing systems is that such systems are hierarchically and modularly constructed from a number of components, which may interact in parallel and which can be viewed as information processing systems themselves. In this case, we call the system a *distributed system*. Distribution here means *spatial distribution* as well as *logical distribution* of functionality across components. However, there are systems which are not parallelized or distributed any further. Such *basic components* can be modeled using state automata with input and output. The repeated decomposition of a system into subsystems yields a hierarchical system, the structure of which can be viewed as a tree with distributed systems on the inner nodes and with basic components on the leaves.

We are interested in a system model in which each kind of interaction is expressible. In our opinion, each kind of interaction can be viewed as the *exchange of a message* between the interacting components. Thus components can be modeled as having *input ports* to receive messages from their environment, and *output ports* to send messages to their environment. The ports constitute the *interface* of a component, they provide the only possibility for the interaction between a component and its environment. The *behavior* of such a component is the relationship between the sequences of messages on its inputs ports and the sequences of messages on its output ports. Systems and their components encapsulate data as well as process. *Encapsulation of data* means that the state is not directly visible to the environment, but can only be accessed using explicit communication. *Encapsulation of a process* means that the exchange of a message does not imply the exchange of control, and that therefore each component is a process of its own.

Exchange of messages between the components of a system is *asynchronous*. This means that a message can be sent independently of the actual readiness of the receiver to receive the message. The requirement for asynchronous communication results from experience in the project FOCUS [BDD$^+$93]: Asynchronous system models provide the most abstract system model for systems with message exchange. They can easily be modeled using stream processing functions, for which a multitude of tractable specification techniques for untimed as well as for timed systems exist ([GS95], [BDD$^+$93]). Moreover, for stream processing functions a powerful theory for compositional refinement has been developed. By using an asynchronous system model, in contrast to process algebraic approaches like the $\pi$-calculus [Mil91] or CCS [Mil89], we do not have to tackle synchronization issues. To take into account synchronization aspects is in our opinion an issue which is irrelevant in the early phases of system development. However, synchronization can easily be encoded in our model, for instance by using an appropriate protocol.

If possible, the system model should not impose any constraints concerning the *addressing of messages*. One possibility for the addressing is that the input- and output ports are statically connected through *channels*. Alternatively, it is also possible in our model to address messages using *identifiers*, as they are used in the context of object-oriented programming languages. Moreover, in defining the semantics of object-oriented programming languages we cannot assume that the set of components is static, but we have to allow for the dynamic generation of components. These requirements lead to two concepts for communication. The first uses ports and the second uses identifiers. The system model has to be prepared for both communication concepts, where one of them or a combination of both may be chosen if the system model is applied. However, our system model is not concerned with further object oriented concepts like class descriptions or inheritance hierarchies. These are regarded as description techniques, the semantics of which is defined using the mathematical system model.

To allow for the consideration of systems in which quantitative time is relevant, the system model has to provide an explicit notion of time which goes beyond the causality relation formalized by the monotonicity requirement for stream processing functions [BDD+93]. We assume that a *discrete time*, which is obtained by partitioning the time scale into equidistant time intervals, is sufficient for the purpose of SYSLAB.

The system model is a reference model, which is referred to by the SYSLAB method description, by the definitions of the semantics of the description techniques, and by the tool development. It serves primarily as a basis for the communication among the people involved in the project, and it has to be presented accordingly. Because issues concerning refinement and verification, as they are treated in the projects FOCUS [BDD+93] and SPECTRUM [BFG+93], play a subordinated role - at least for the present - it is not necessary to provide a concrete syntax or a deduction calculus for the system model, or to code the system model in a formal logic. Therefore, we restrict ourselves to a purely mathematical presentation of the system model. However, it is possible that future enhancements of the system model will obtain a more formal syntax and semantics.

This paper is organized as follows: In the next section the black-box view of systems is presented. This is done by describing the mathematical structure of streams, by presenting stream processing functions as a model of interactive systems, and by introducing identifiers for components. In section 3 we introduce two glass-box views, the system as a *basic component* and the system as a *distributed system*. In section 4 we give a conclusion by comparing the presented system model with the requirements stated in this section.

# 2  Black-Box View

An information processing system is an entity interacting with its environment by the exchange of messages. The interface between the system and the environment can be modeled as consisting of so called *ports*, which are often also called *channels*, over which data flow. We distinguish between *input ports* and *output ports*. A graphical representation of a component with the input ports $port_1$ and $port_2$ and the output ports $port_3$, $port_4$ and $port_5$ is given in Figure 1. We assume that all port names like $port_1 \ldots port_5$ are contained in the set $P$ of *port names*, which is required to be at most countable.

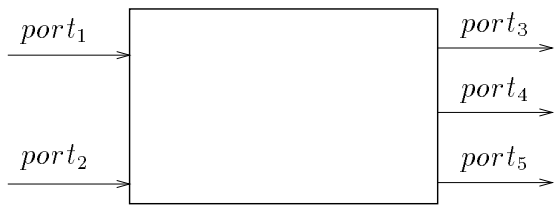

Figure 1: Black-box view of a system

At runtime, a system receives messages on its input ports and sends messages on its output ports according to its behavior. In the sequel, we will start by introducing *streams* as a model for the communication history of ports, after which we present *stream processing functions* as a model of interactive systems and *identifiers* of components in our system model.

## 2.1  Streams

The behavior of a system is modeled by its system runs, which describe the relationship between the messages arriving on the input ports of the system and the messages sent on the output ports of the system. We assume that for each run the events on a port are totally ordered, which means that for two different events always one causally and temporarily precedes the other. This allows to model the communication history on a port by a stream of messages.

A *stream* is a finite or infinite sequences of messages. If $M$ denotes the set of messages, $M^*$ the set of all finite sequences of messages and $M^\infty$ the set of all infinite sequences of messages, for the set of all streams over $M$, denoted by $M^\omega$, we can define:

$$M^\omega = M^\infty \cup M^*$$

We will use the following operations on streams:

- $\widehat{\phantom{x}} : M^\omega \times M^\omega \to M^\omega$ denotes the concatenation of two streams. Thus $s\widehat{\phantom{x}}t$ is the stream which is obtained by putting the second argument after the first. The operator $\widehat{\phantom{x}}$ is usually written in infix notation. We assume that

  $$s \in M^\infty \Rightarrow s\widehat{\phantom{x}}t = s,$$

  holds, which states that the concatenation of an infinite stream $s$ with a stream $t$ yields the stream $s$. $\widehat{\phantom{x}}$ will also be used to concatenate a single message with a stream.

- $\# : M^\omega \to \mathbb{N} \cup \{\infty\}$ delivers the length of the stream as a natural number or $\infty$, if the stream is infinite.

- $Filter : \mathcal{P}(M) \times M^\omega \to M^\omega$ denotes the filter-function. $Filter(N, s)$ deletes all elements in $s$ which are not contained in set $N$.

In addition to the total order of events modeled by the data-type of streams our system model also provides an explicit notion of time. Like in [Stø95] we assume that time proceeds in equidistant time intervals, and we model the proceeding of time by one time interval using a time signal $\sqrt{\phantom{x}} \notin M$, called *tick*. With $M^{\sqrt{\phantom{x}}}$ we denote the set $M \cup \{\sqrt{\phantom{x}}\}$, and we define:

$$M^{\overline{\infty}} = \{s \in (M^{\sqrt{\phantom{x}}})^\omega \mid \#(Filter(\{\sqrt{\phantom{x}}\}, s)) = \infty\}$$
$$M^{\overline{*}} = (M^{\sqrt{\phantom{x}}})^*$$

The set $M^{\overline{\infty}}$ is the set of all infinite sequences of elements from $M \cup \{\sqrt{\phantom{x}}\}$, which contain infinitely many copies of $\sqrt{\phantom{x}}$. The requirement for infinitely many copies of $\sqrt{\phantom{x}}$ models the fact that time never ends and that we consider only infinite communication histories. Streams over $M^{\sqrt{\phantom{x}}}$ contain only finitely many messages from $M$ between two ticks. The set $M^{\overline{*}}$ will be used in the sequel to speak about finite prefixes of infinite streams.

Assuming that *In* denotes the set of all input ports and that *Out* denotes the set of all output ports, the communication history of a system can be modeled by a pair of functions *in* and *out*, which map ports to streams of messages and ticks:

$$in : In \to M^{\overline{\infty}}$$
$$out : Out \to M^{\overline{\infty}}$$

Functions like *in* and *out*, which map port names to timed streams, are called *bunches* of message streams. This way, the selection of a message stream of port $p$ out of a bunch of messages $b$ corresponds to function application. To ease readability, in this case we write the function application in the form

$$b.x$$

where $x \in In \cup Out$.

## 2.2 Stream processing functions

The *behavior* of a system is modeled by a *timed stream processing function* mapping a bunch of input streams to a bunch of output streams:

$$Behavior : (In \to M^{\overline{\infty}}) \to (Out \to M^{\overline{\infty}})$$

However, not every function with this functionality represents an adequate model of an information processing system: In reality, it is impossible that at any point of time the output depends on future input. To model this fact, we impose an additional mathematical requirement. First we define:

$$\downarrow : M^{\overline{\infty}} \times Nat \to M^{\overline{*}}.$$

The application of $\downarrow$ will be written in infix notation. $s \downarrow j$ yields the first $j$ time intervals of the stream $s$, i.e. $s \downarrow j$ is the prefix of $s$ containing the $j$.th tick as last element, or the empty stream if $j = 0$. For that reason, $s \downarrow j$ contains exactly $j$ ticks, and $s \downarrow j$ is a prefix of $s$:

$$\#(\mathit{Filter}(\{\sqrt{}\}, s \downarrow j)) = j$$
$$\exists t : M^{\overline{\infty}} : (s \downarrow j)\hat{\ }t = s$$
$$j = 0 \Rightarrow \#(s \downarrow j) = 0$$
$$j > 0 \Rightarrow \exists t : M^{\overline{*}} : s \downarrow j = t\hat{\ }\sqrt{}$$

The operator $\downarrow$ is overloaded to bunches of infinite timed streams by point-wise application. Let $s : L \to M^{\overline{\infty}}$ with $L \subseteq P$ be such a bunch of timed streams:

$$(s \downarrow j).p = (s.p) \downarrow j$$

We now postulate the requirement that the output of a component at any point of time $j$ may not depend on the input at a future point of time. This would result in an oracle, which is not implementable. We therefore require stream processing functions to be *pulse-driven*. The function $Behavior$ is called *pulse-driven*, if for each $j$, the output up to to time $j$ is only determined by the input up to time $j$:

$$s \downarrow j = t \downarrow j \Rightarrow Behavior(s) \downarrow j = Behavior(t) \downarrow j$$

Functions with a bunch of input streams as domain and a bunch of output streams as range that are pulse-driven are called *stream processing functions*. We denote the set of stream processing functions by

$$(In \to M^{\overline{\infty}}) \xrightarrow{p} (Out \to M^{\overline{\infty}}).$$

To use stream processing functions to model behavior of systems gives us a very simple composition technique for components, based on function composition.

In the following, we characterize the set of all distributed systems we are interested to model. This is done by characterizing properties of all instances of the system model.

## 2.3 Identifiers

We are interested in systems that allow to address a message by the *identifier* of the receiver, like this is in general done in object-oriented programming languages. We use a countable set $ID$ of identifiers for this purpose. Every identifier names exactly one component in the system and every component has exactly one identifier. However every component may have several input and output ports. We denote them by functions $In_{id}$ and $Out_{id}$, that attach sets of portnames to every identifier:

$$In : ID \to \mathcal{P}(P)$$
$$Out : ID \to \mathcal{P}(P)$$

The application of $In$ and $Out$ is written as $In_{id}$ and $Out_{id}$. We require the sets of portnames of different components to be disjoint:

$$id \neq id' \Rightarrow (In_{id} \cup Out_{id}) \cap (In_{id'} \cup Out_{id'}) = \emptyset$$

This requirement does not restrict the power of our system model, but simplifies the definitions in the sequel, because now every portname is uniquely attached to one component.

Identifiers and portnames serve two purposes. On one hand, they allow us to model components resp. channels during the system development, on the other hand, they can be used for the implementation of message passing mechanisms. In the second case identifiers or portnames become part of the messages which flow within the system.

A stream processing function that models the behavior of a system component with identifier $id$ is denoted as

$$Behavior_{id} : (In_{id} \to M^{\overline{\infty}}) \xrightarrow{p} (Out_{id} \to M^{\overline{\infty}}).$$

Function $Behavior_{id}$ exactly describes the result on the output ports for every input given on the input ports.

## 3 Glass-Box Views

As already mentioned in the beginning, regarding the internal construction, we distinguish between

- basic components and
- distributed systems that are decomposed into a nonempty set of components.

The set of identifiers $ID$ can therefore be divided into the disjoint sets of identifiers for basic components $ID_b$ and of identifiers for distributed components $ID_s$:

$ID = ID_b \cup ID_s$
$ID_b \cap ID_s = \emptyset$

### 3.1 Basic components

Basic components are systems that are not composed of distributed components. They can be modeled by stream processing functions or by state-machines with input and output. Mathematical models for basic components are for example state-transition-systems [BDDW91] or I/O-automata [LS89]. Especially concurrent timed port automata [GR95] are suited to describe basic components with several input and output ports in a timed environment.

A description of basic components by state-machines is suitable whenever concrete assumptions about the structure of the internal state of the component are made. If a description-technique only considers the black-box behavior of a component, we will not explicitly construct state-machines, but instead we will use a characterization of the behavior just by stream processing functions.

## 3.2 Distributed Systems

Besides being a basic component, a component can internally be decomposed into a set subsystems called *components*. In this case we speak of a *distributed system*. As already mentioned, distribution in this case means spatial distribution as well as logical distribution. The identifiers of the components of a distributed system are denoted by *Parts*:

$$Parts : ID_s \to \mathcal{P}(ID)$$

By repeated decomposition of a system we get a hierarchy of systems and subsystems. Function *Parts* therefore characterizes a tree, with a special identifier

$$RootSystem \in ID$$

as root of this tree. By this arrangement of all components in a component hierarchy, the superior components as well as the parts of a component are uniquely determined. The set of identifiers together with function *Parts* is used to define this hierarchical structure of systems, while the set of portnames determines communication channels.

We now examine the relationship between the behavior of a distributed system $id \in ID_s$ and the behaviors of its components. By $InParts_{id}$ and $OutParts_{id}$, we denote the sets of input and output ports of all components of *id*. They are defined as follows:

$$InParts_{id} = \{p | \exists id' \in Parts(id) : p \in In_{id'}\}$$
$$OutParts_{id} = \{p | \exists id' \in Parts(id) : p \in Out_{id'}\}$$

Figure 2 shows a diagram of a distributed system. A distributed system consists of its components *Parts(id)* and a *communication medium*, which transmits the messages from the sender to the correct port of the receiver. The communication medium acts like a "membrane" between the inner and the environment of a component. In the following, we characterize the message flow through this membrane by relating the input and the output message streams of this membrane.

## 3.3 The Communication Medium

The communication medium has a complex signature, the *message origins* $Origins_{id}$ and the *message destinations* $Destinations_{id}$. The message origins consist of the input ports of system *id* and of the output ports of the components of *id*. Conversely the message destinations consist of the output ports of *id* and of the input ports of the components of *id*.

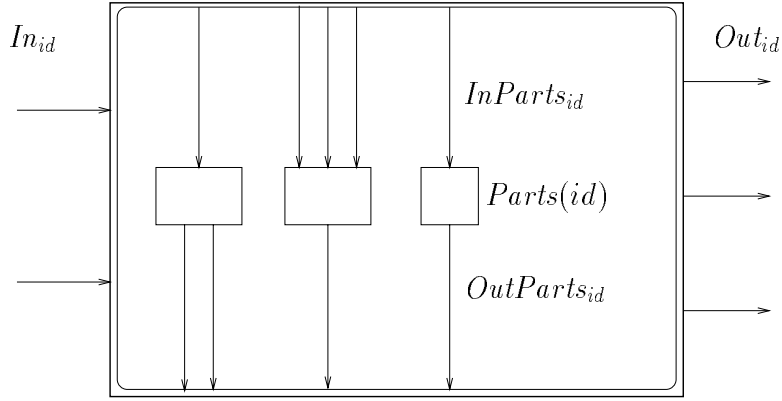

Figure 2: A distributed system

$Origins_{id} = In_{id} \cup OutParts_{id}$
$Destinations_{id} = Out_{id} \cup InParts_{id}$

For description purposes, we assume that every message contains its origin and destinations in itself. We therefore do not allow message broadcasting, but require that every message carries the information that identifies a unique destination. We model this by two functions

$origin_{id} : M \to Origins_{id}$
$destination_{id} : M \to Destinations_{id},$

that describe the origin and the destination port of a message depending on the system *id* through which the message actually flows. The two functions $origin_{id}$ and $destination_{id}$ define the connection structure between the components of a distributed system. If we have an object-oriented system, messages carry their destination identifier and $destination_{id}$ solely depends on this identifier. If we have hard-wired systems, such as hardware systems, function $destination_{id}$ may only depend on function $origin_{id}$, where it is required that messages with the same origin have the same destination.

We require that the following properties w.r.t the message flow hold within the system model:

- For each input port of the system and for each output port of a component the order of messages sent to a certain destination has to be maintained. This requirement enforces a linear ordering of messages within every connection.

- The contents of messages may not be modified. Messages cannot be duplicated or lost. No new messages are generated.

A lot of systems exhibit connection structures where these requirements for message transmission are not valid. These systems can easily be encoded within our system model if we use special transmitter components exhibiting the behavior of such a connection structure.

We do not require our communication medium to be free of *delay*, since we do not impose any requirement on the time difference between the sending and the receiving of a message besides the requirement that this time is finite.

We are now able to specify a communication medium that distributes messages according to the above requirements by relating origin and destination streams of the communication medium. Let

$$ostreams : Origins_{id} \to M^{\overline{\infty}}$$
$$dstreams : Destinations_{id} \to M^{\overline{\infty}}$$

be timed streams of messages for the input and output ports. Then we have:

1. Origin and destination streams restricted to the input resp. output ports of system $id$ exhibit the behavior of system $id$:

   $$Behavior_{id}(ostreams|_{In_{id}}) = dstreams|_{Out_{id}}$$

   With $f|_M$ we denote the restriction of a function $f : N \to L$ with $M \subseteq N$ to set $M$. Therefore the restriction $ostreams|_{In_{id}}$ selects the bunch of streams that flow on the input ports of the system only. Accordingly $dstreams|_{Out_{id}}$ selects the streams on the output ports of the system.

2. Input and output ports of every component have to exhibit message streams according to their behavior:

   $$id' \in Parts(id) \Rightarrow Behavior_{id'}(dstreams|In_{id'}) = ostreams|Out_{id'}$$

3. Every destination stream actually contains the messages for this destination port:

   $$Filter(\{m|origin_{id}(m) = s\} \cup \{\sqrt{}\}, dstreams.d)$$
   $$= Delay(Filter(\{m|destination_{id}(m) = d\} \cup \{\sqrt{}\}, ostreams.s))$$

If the message stream of destination port *dstreams.d* is filtered for messages coming from origin port *s*, we get a similar message stream, as if we filter the messages of origin stream *ostreams.s* for messages to destination port *d*. The only difference is possible delay of messages, but no rearrangement of ordering, duplication or loss of messages. Delay is modeled by the following pulse-driven stream processing function:

$$Delay : M^{\overline{\infty}} \xrightarrow{p} M^{\overline{\infty}}$$
$$Filter(M, Delay(s)) = Filter(M, s)$$

From the definition of pulse-driven stream processing functions, it follows that *Delay* really delays messages.

# 4 Discussion and Concluding Remarks

In this paper a so-called system model has been presented as an abstract mathematical model for information processing systems. Because the model is based on FOCUS [BDD+93], a mathematical modeling and development technique for distributed systems, a multitude of refinement and verification techniques for the system model exists. The presented model allows for the formal foundation and semantic integration of a large class of description- and programming techniques. The applicability ranges from analysis, specification and design documents to programs in (distributed) object-oriented programming languages. An explicit notion of time makes the model also well-suited for real-time and hardware systems. The flexibility of the system model is to a large extent possible due to the underspecification of the communication medium which allows for a large number of different applications.

A lot of open problems are to be tackled with this model. First of all, dynamic creation of components exists only implicitly. A component that starts to act only if it gets an initial creation message may be regarded as a component which is not created until the creation message arrives. Similarly deletion of components may be encoded. Only experience will show whether this is tedious, when proving properties of systems. Another problem is that it is lengthy and to some extent intricate to model systems directly within this system model. Instead we propose a coherent set of description techniques, that do not only exhibit a formal syntax, but also a formal semantics based on the system model. This is done within the SYSLAB project, for which the system model is a vital part.

# References

[AG90]      C. Ashworth and M. Goodland. *SSADM: A Practical Approach.* McGraw-Hill, 1990.

[BDD+93]    M. Broy, F. Dederichs, C. Dendorfer, M. Fuchs, T.F. Gritzner, and R. Weber. The Design of Distributed Systems - An Introduction to FOCUS. Technical Report SFB 342/2/92 A, Technische Universität München, Institut für Informatik, 1993.

[BDDW91]    M. Broy, F. Dederichs, C. Dendorfer, and R. Weber. Characterizing the behaviour of reactive systems by trace sets. Technical Report SFB 342/2/91 A, TUM-I9102, Technische Universität München, Institut für Informatik, February 1991.

[BFG+93]    M. Broy, C. Facchi, R. Grosu, R. Hettler, H. Hussmann, D. Nazareth, F. Regensburger, O. Slotosch, and K. Stølen. The Requirement and Design Specification Language SPECTRUM, An Informal Introduction, Version 1.0. Technical Report TUM-I9311, Technische Universität München, Institut für Informatik, May 1993.

[CAB+94]    D. Coleman, P. Arnold, S. Bodoff, C. Dollin, H. Gilchrist, F. Hayes, and P. Jeremaes. *Object-Oriented Development - The Fusion Method.* Prentice Hall, Inc., Englewood Cliffs, New Jersey, 1994.

[CDI92]     Introduction to CDIF: The CASE data interchange format standards, April 1992. CDIF Technical Commitee, LBMS, Evelyn House, 62 Oxford Street, London W1N 9LF, United Kingdom.

[GR95]      R. Grosu and B. Rumpe. Concurrent timed port automata. Technical Report TUM-I 9533, Technische Universität München, 1995.

[GS95]      R. Grosu and K. Stølen. A Denotational Model for Mobile Point-to-Point Dataflow Networks. Technical Report TUM-I 9527, Technische Universität München, 1995.

[Hel90]     G. Held, editor. *Sprachbeschreibung GRAPES.* SNI AG, München Paderborn, 1990.

[HL93]      H.J. Habermann and F. Leymann. *Repository - eine Einführung.* Handbuch der Informatik. Oldenbourg, 1993.

[Hus94]     H. Hussmann. Formal Foundations for SSADM: An Approach Integrating the Formal and Pragmatic Worlds of Requirements Engineering. Habilitation thesis, Technische Universität München, July 1994.

[LS89]     N. Lynch and E. Stark. A proof of the kahn principle for input/output automata. *Information and Computation*, 82:81–92, 1989.

[Mil89]    R. Milner. *Communication and Concurrency*. Prentice Hall, Englewood Cliff, 1989.

[Mil91]    R. Milner. The polyadic $\pi$-calculus: A tutorial. Technical Report ECS-LFCS-91-180, October 1991.

[RBP+92]   J. Rumbaugh, M. Blaha, W. Premerlani, F. Eddy, and W. Lorensen. *Object-Oriented Modeling and Design*. Prentice Hall, Inc., Englewood Cliffs, New Jersey, 1992.

[SFD92]    L.T. Semmens, R.B. France, and T.W.G. Docker. Integrated Structured Analysis and Formal Specification Techniques. *The Computer Journal*, 35(6):600–610, 1992.

[Stø95]    K. Stølen. A Framework for the Specification and Development of Reactive Systems. Draft, 1995.

[Tho89]    I. Thomas. PCTE interfaces: Supporting tools in software-engineering environments. *IEEE Software*, pages 15–23, November 1989.